\documentclass[12pt]{article}
\usepackage{graphicx}
\usepackage{amssymb}

\def\vv{\vspace{-1ex}}

\begin{document}

\begin{center}

{\Large
A SiPM-based ZnS:$^6$LiF scintillation neutron detector
}

\vspace{2ex}
A.\,Stoykov, J.-B.\,Mosset, U.\,Greuter, M.\,Hildebrandt, N.\,Schlumpf

\vspace{2ex}

Paul Scherrer Institut, CH-5232 Villigen PSI, Switzerland \\[1ex]

\end{center}

\vspace{2ex}
\noindent
In the work presented here we built and evaluated a single-channel neutron detection unit
consisting of a ZnS:$^6$LiF scintillator with embedded WLS fibers readout by a SiPM.
The unit has a sensitive volume of 2.4\,x\,2.8\,x\,50\,mm$^3$;
12 WLS fibers of diameter 0.25\,mm are uniformly distributed over this volume
and are coupled to a 1\,x\,1\,mm$^2$ active area SiPM.
We report the following performance parameters:
neutron detection efficiency $\sim 65$\,\% at 1.2\,\AA,
background count rate $< 10^{-3}$\,Hz,
gamma-sensitivity with $^{60}$Co source $< 10^{-6}$,
dead time $\sim 20\,\mu$s, multi-count ratio $< 1$\,\%.
All these parameters were achieved up to the SiPM dark count rate of $\sim 2$\,MHz.

We consider such detection unit as an elementary building block for realization
of one-dimensional multichannel detectors for applications in the neutron scattering
experimental technique. The dimensions of the unit and the number of embedded fibers
can be varied to meet the specific application requirements.
The upper limit of $\sim 2$\,MHz on the SiPM dark count rate allows
to use SiPMs with larger active areas if required.

\vspace{3ex}

\vspace{2ex}
\section{Introduction}
Helium-3 has been for several decades the most widely used converting material
in detectors for neutron scattering experiments. The world-wide shortage of its supply
starting in 2009 increased significance and stimulated further development of
alternative detector technologies \cite{1}. One of these alternatives is
the scintillation technology based on ZnS:$^6$LiF or ZnS:$^{10}$B$_2$0$_3$ scintillators
read out by wavelength-shifting (WLS) fibers \cite{1,2}.
Currently all detectors of this kind utilize photomultiplier tubes (PMTs)
or multi-anode photomultiplier tubes (MaPMTs) as photosensors.

The application of silicon photomultipliers (SiPMs) in such detectors has been hindered
by their orders of magnitude higher dark count rate at room temperature:
the long emission time of the neutron scintillator and the deficient light collection
due to its poor transparency made it difficult to combine a high trigger efficiency
for the neutron signals with a reasonable suppression of the SiPM dark counts.
As mentioned in \cite{2}, the solution of this problem requires an improvement
of the light collection from the scintillator.
In \cite{3,4,5} we presented a practical way how to combine this requirement with
a small active area of a SiPM. Also we developed an approach to the signal processing
based on ``digitization'' of the SiPM one-electron signals
(one primary electron = one standard pulse) followed by a pulse-train analysis
to identify the neutron related pulse sequences against the background
of the SiPM dark counts.

In this work we built a single-channel detection unit with SiPM readout
(prototype units of 1/4 height were used in \cite{3,4,5})
and characterized its performance by determining such parameters as trigger efficiency,
background count rate, gamma-sensitivity, dead time, and multi-count ratio.

\section{Detection unit}
Figure~1 shows a cross-section of the sensitive volume of the detection unit.
The unit consists of four times two layers (thickness 0.25\,mm and 0.45\,mm)
of ZnS:$^6$LiF scintillation material
(ND2:1 neutron detection screens from Applied Scintillation Technologies \cite{6})
glued together using EJ-500 optical epoxy from Eljen \cite{7}.
Twelve WLS fibers Y11(400)M from Kuraray \cite{8} are glued with the same epoxy
into the grooves machined in the thicker layers.
Compared to \cite{3,4,5} we use here WLS fibers with twice higher dye concentration
(400\,ppm instead of 200\,ppm). This increases the light yield by about 20\,\%.
At one side of the unit the fibers are cut along its edge and polished.
An aluminized Mylar foil serving as specular reflector is glued on these polished fiber
ends to increase the light yield at the other fiber ends which are connected to a SiPM.
The total volume of the unit is 2.4\,x\,2.8\,x\,50\,mm$^3$ (width x height x length).
The net absorption volume excluding grooves with the fibers (effective volume)
amounts to 0.83 of this value.

The free ends of the WLS fibers are bundled and glued together into $\oslash 1.1$\,mm hole
in a Plexiglas holder and polished afterwards. The coupling to a 1\,x\,1\,mm$^2$
active area SiPM is done via a so-called optical expander
(short $\oslash 1.2$\,mm clear multiclad fiber) to ensure a uniform illumination
of the SiPM sensitive area. Note that the scintillation light from a neutron absorption
event is not distributed uniformly over the 12 WLS fibers but is rather concentrated
in one or few of them.

The width and the height of this detection unit satisfy the requirements of the POLDI
time-of-flight diffractometer concerning the channel pitch (2.5\,mm) and the neutron
absorption probability ($\sim 80$\,\% at 1.2\,\AA) \cite{5}.
A one-dimensional array of 400 such units of 200\,mm length arranged along a circle
of 2\,m radius will constitute one detector module of the POLDI instrument
(in total the instrument will be equipped with four such modules).
The 50\,mm length of the unit in the present work is chosen arbitrarily:
we do not expect variations in the performance of the detector and complications
in the manufacturing process changing later to the length of 200\,mm.

\section{Signal processing}
The used SiPM is a 1\,x\,1\,mm$^2$ active area MPPC S12571-025C from Hamamatsu \cite{9}.
It is operated at an overvoltage of 2.5\,V at room temperature. The dark count rate is
about 100\,kHz. Higher dark count rates were induced by illuminating the SiPM with
a weak constant light source. The necessity to evaluate the performance of the detector
at the dark count rates substantially higher than 100\,kHz is motivated by the following
considerations. First, even though 1\,mm$^2$ active area SiPMs with such low dark count
rates start to be commonly available, our goal is to develop a detector which will fulfil
the performance requirements with different types of SiPMs,
e.g. devices with higher intrinsic dark count rates or larger active areas.
And second, depending on the radiation environment, the dark count rate of the SiPM
might increase with time conditioned by an increase of the concentration of radiation
defects in silicon \cite{10}. For example, indicated by long-term measurements
in the POLDI experimental area we expect an increase of the SiPM dark count rate
of about 100\,kHz/mm$^2$ per year.

Figure~2 shows the block-diagram of the signal-processing chain including
a high band-width amplifier, a leading-edge discriminator,
an analyzer (designated here as filter), and an event generator unit.

Figure~3 shows an oscilloscope ``screen-short'' (persistence mode) of the amplified
and shaped SiPM signals (SA) and the generated discriminator signals (SD).
The discrimination threshold is set low enough so that all SA-signals are accepted.
Independent of how many SiPM cells are triggered simultaneously by SiPM cell-to-cell
cross-talk per initial single primary charge carrier in one cell
(see multiple amplitudes of SA signals), after the discriminator the relation
``one primary charge carrier = one standard SD pulse'' is always given.
This kind of ``digitization'' suppresses the cross-talk events, which is essential
to achieve low background count rate of the detector \cite{5}, and allows for further
signal processing schemes independent of the used SiPM type.

The SD pulse sequence is processed by the analyzer unit. Approaches for the realization
of the analyzer can be chosen differently as described in \cite{3,4,5}.
In the current case we use a multistage ``single-pulse elimination'' filter
described in \cite{4} to extract the neutron signals from the dark-count background.
The tunable parameters of the filter, defining the detection threshold,
are the number of filtering stages N and the width of the internal gate signals
for the first and the following filter stages: gate(1) and gate(2$\ldots$N).
The dark count rejection is dominated by the gate width of the first filter stage
and by the total number of consecutive stages, while for better transmission
of the ``extended'' pulse trains corresponding to neutron scintillation events
longer gate values for the following stages are advantageous.
In this study, the filter time constants gate(2$\ldots$N) were fixed to 500\,ns,
while the parameters gate(1) and N were varied.

Figure~4 shows an example of SD pulse sequence and of corresponding SF pulse sequence
from the filter as a result of ``identification'' of one neutron scintillation event.
The first pulse of the SF-signal triggers the event generator
(a retriggerable mono-flop with adjustable pulse width) which generates an event signal SN.
For the duration of the SN signal the system is blocked -- no second pulse can be generated.
Such blocking is necessary to account for the long afterglow of the scintillator
and prevents multiple triggers from the same scintillation event. The actual blocking time
is automatically adjusted in accordance with the strength of the signal at the filter
output and is equal or longer than its initial set value (b-time). The parameter b-time
was varied in the present measurements in the range from 5\,$\mu$s to 200\,$\mu$s.

\section{Measurements}

\subsection{Trigger Efficiency}
The measurements were performed with a $^{241}$AmBe neutron source
(intensity $2\cdot10^4$ fast neutrons per second) positioned in the center of
a 0.8\,x\,0.8\,x\,0.8\,m$^3$ moderator block made of polyethylene.
The detection unit was placed in a distance of $\sim 5$\,cm from the source in
a slot made in the moderator block. To shield the detector from 60\,keV gamma photons
accompanying the alpha-decay of $^{241}$Am, the source was shielded by placing it
inside a cylindrical tube made of lead with 5\,mm thick walls.

To estimate the neutron absorption rate in the detection unit a calibration measurement
with a low threshold (gate(1)\,=\,80\,ns, N\,=\,4) was performed.
Figure~5 shows the distribution of the number of SD-counts in the first 10\,$\mu$s
of the signals. The trigger efficiency of 0.92 is obtained as the ratio of the number
of events in this measured histogram to that in the histogram obtained by extrapolating
the measured histogram to zero number of SD-counts. Taking into account the measured
event rate of 8.3\,Hz the neutron absorption rate in the detection unit is estimated
to be 9.0\,Hz. In the following measurements the trigger efficiency was determined
as the ratio of the measured event rate to this neutron absorption rate.

Figure~6 shows the trigger efficiency as a function of the filter parameters
gate(1) and N. By decreasing gate(1) and increasing N we increase the strength
of the filter and thus decrease its trigger efficiency. The operation point was chosen
at gate(1)\,=\,30\,ns and N\,=\,10 which ensures the trigger efficiency at the level
of 80\,\%. Combined with the neutron absorption probability of $\sim 80$\,\%
at 1.2\,\AA (see above) this gives the neutron detection efficiency of
$\sim 65$\,\% at this wavelength.

\subsection{Background Count Rate}
Figure~7 shows the background count rate of the detector as a function of the number
of filter stages for three different values (100~kHz, 1000~kHz, and 2000~kHz)
of SiPM dark count rates. At the chosen operation point, where we reach
a trigger efficiency around 80\,\%, the background count rate below 10$^{-3}$\,Hz
is achieved for SiPM dark count rates of up to $\sim 2$\,MHz.

\subsection{Gamma Sensitivity}
The gamma-sensitivity was measured with a $^{60}$Co source. The source is point-like
and incorporated into a tablet of $\oslash = 25$\,mm. Its activity is $\sim 60$\,kBq.
The rate of $\sim 1.3$\,MeV photons passing through our 2.4\,mm wide detection unit
was estimated to be $10^4$\,s$^{-1}$.
This calibration was done with a 2\,x\,2\,x\,12\,mm$^3$ LYSO crystal mounted onto
a photocathode of a PMT: with the low detection threshold the rate of gamma-events
was measured to be $\sim 10^3$\,s$^{-1}$ and the interaction probability
for 1.3\,MeV photons in 2\,mm thick LYSO material was taken as 10\,\% \cite{11}.

Figure~8 shows the gamma-sensitivity as a function of the number of filter stages
at different SiPM dark count rates. The probability to trigger on gamma-events
is enhanced by the presence of dark counts and the gamma-sensitivity increases
with increasing the dark count rate. At the chosen operation point the gamma-sensitivity
amounts to $3\cdot10^{-8}$, $10^{-7}$, and $8\cdot10^{-7}$ at the SiPM dark count rates
of 140\,kHz, 1000\,kHz, and 2000\,kHz, respectively.

\subsection{Multi-Count Ratio}
Figure~9 shows the multi-count ratio of the detector as a function of the set value
of the blocking time (b-time) of the event generator.
The multi-count ratio is defined \cite{12} as the ratio between the measured
and the true number of neutron events minus 1. For the true number of neutron events
we take the value measured with b-time\,=\,200\,$\mu$s. Down to b-time\,=\,15\,$\mu$s
the multi-count ratio remains zero within the experimental accuracy of $\sim 0.05$\,\%.
Below 15\,$\mu$s the multi-count ratio starts to increase and reaches the level of 1\,\%
at b-time~$\sim 8\,\mu$s. At b-time\,=\,15\,$\mu$s (chosen setting)
the actual measured mean value of the blocking time (dead time) is $\sim 20\,\mu$s.

\subsection{Influence of SiPM dark count rate}
Figure~10 shows how the trigger efficiency and the multi-count ratio of the detector
are influenced by the dark count rate of the SiPM. The trigger probability is enhanced
by the presence of dark counts and both the trigger efficiency and the multi-count ratio
increase with increasing the dark count rate. However, up to a dark count rate
of $\sim 2$\,MHz the variation of these two parameters is minor.

\section*{Summary}
In this work we continued investigating the feasibility of the new approach,
proposed in \cite{3,4,5}, to realize one-dimensional multichannel neutron detectors
with ZnS:$^6$LiF scintillators. In this approach the detector is built as an array
of single-channel detection units containing scintillator with embedded WLS fibers
readout by SiPMs. We built a prototype detection unit and performed characterization
studies demonstrating the performance parameters to be similar to those of the currently
used PMT-based detection systems \cite{12,13}.

It is worth pointing out, that the compact size and the low price of SiPMs make
the discussed approach feasible even in case of certain 2D-detectors.
The large number of individual readout channels in such detectors will help solving
the problem of their limited rate capability (remember, that the count rate per readout
channel is limited by the necessity to introduce a certain dead-time
in the detection process to cope with the long afterglow of the scintillator).
This could be an option, for example, for large-area detectors for inelastic neutron
scattering instruments where currently no adequate replacement to helium-3 detectors
is found \cite{1}.

\section*{Acknowledgments}
We express our gratitude to Andreas Hofer (Detector Group of the Laboratory
for Particle Physics) for designing and building our prototype detection units.

\clearpage
\newpage
\begin{figure}[t]
\centering
\includegraphics[width=0.65\columnwidth,clip]{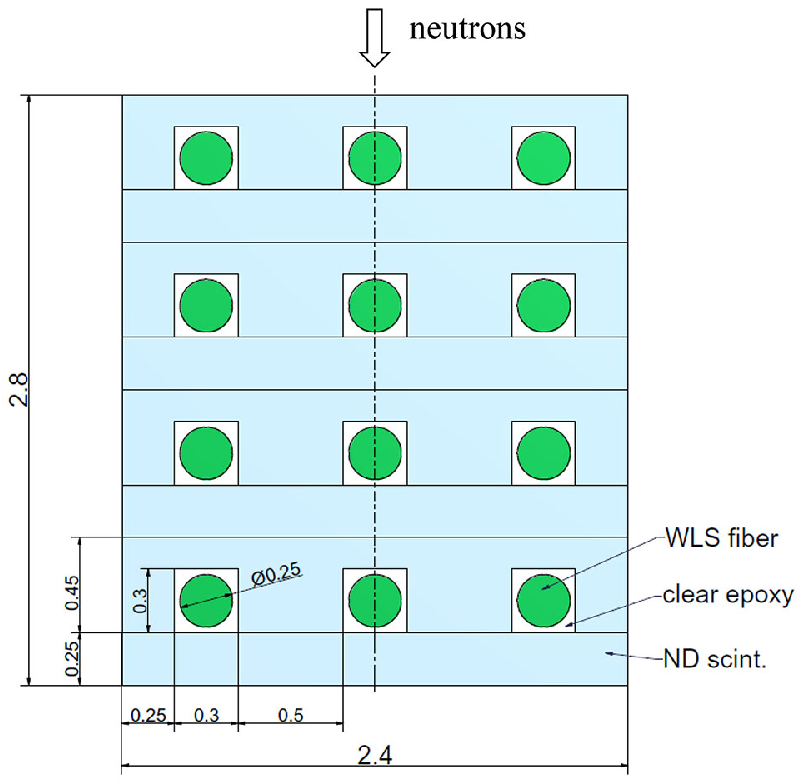}
\caption{
Cross-section of the sensitive volume of the detection unit. The scintillators are
ND2:1 neutron detection screens from Applied Scintillation Technologies, the WLS fibers
are of type Y11(400)M from Kuraray. The optical epoxy used to glue the fibers into
the grooves and the scintillator sheets together is EJ-500 from Eljen.
}
\end{figure}

\begin{figure}[b]
\centering
\includegraphics[width=1.0\columnwidth,clip]{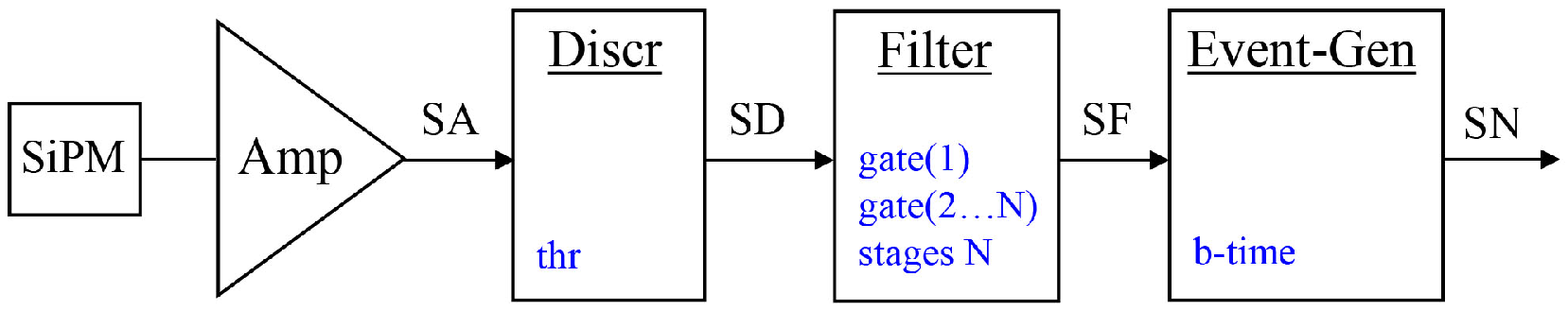}
\caption{
Processing scheme of the SiPM signals: high band-width amplifier,
leading-edge discriminator, multistage filter, and event generator unit.
The tunable parameters are: the discrimination threshold (thr),
the number of filter stages (N), the retriggerable gate width of the first gate(1)
and the following gate(2$\ldots$N) filter stages, and the duration of the output pulse
of the event generator (b-time – blocking time) to account for the afterglow photons
from the scintillation process.
}
\end{figure}

\clearpage
\newpage

\begin{figure}[t]
\centering
\includegraphics[width=0.7\columnwidth,clip]{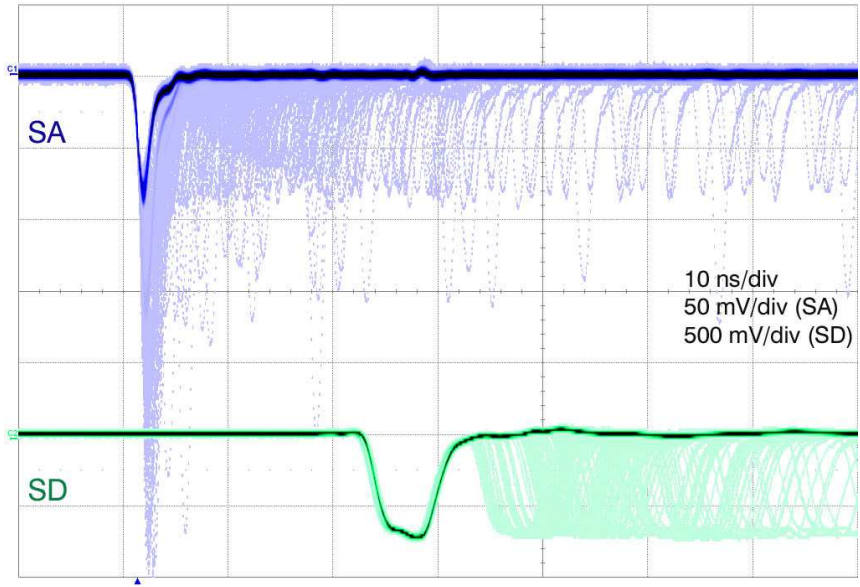}
\caption{
Conversion of the amplified and shaped SiPM analog signals (SA) into standard
digital signals (SD) by a fast leading-edge discrimination stage. Note, that in contrast
to the SA signals having admixture of the SiPM cell-to-cell cross-talk events
(non-zero probability of signals with multiple amplitudes), the SD signals are free
from any cross-talk contribution.
}
\end{figure}

\begin{figure}[b]
\centering
\includegraphics[width=1.0\columnwidth,clip]{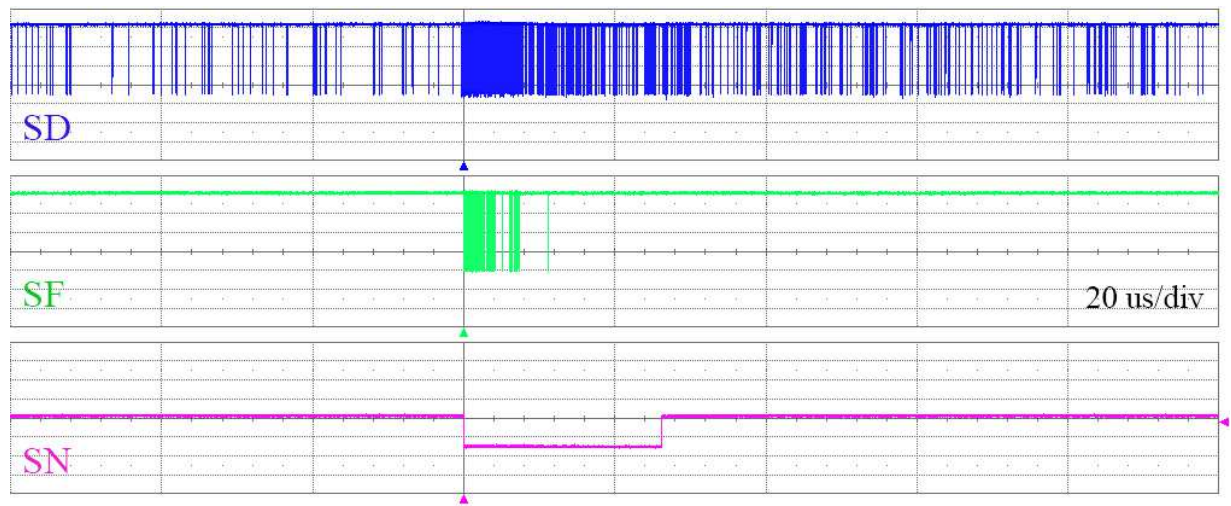}
\caption{
Detection of a neutron event: initial SD pulse sequence and the subsequence
of SD-pulses passing the filter (SF-signal). The first pulse of the SF-sequence
generates the leading edge of an event signal (SN). The set value for the width of
the SN-signal is b-time\,=\,15\,$\mu$s, its actual width determined by the strength
of the SF-signal is $\sim 25\,\mu$s. The presented event is relatively strong:
there are 238 SD-pulses within the first 10\,$\mu$s of the event, 113 of these pulses
pass the filter. The filter settings are: N\,=\,10, gate(1)\,=\,30\,ns,
gate(2$\ldots$N)\,=\,500\,ns. The dark count rate of the SiPM is 1\,MHz.
}
\end{figure}

\clearpage
\newpage

\begin{figure}[t]
\centering
\includegraphics[width=1.0\columnwidth,clip]{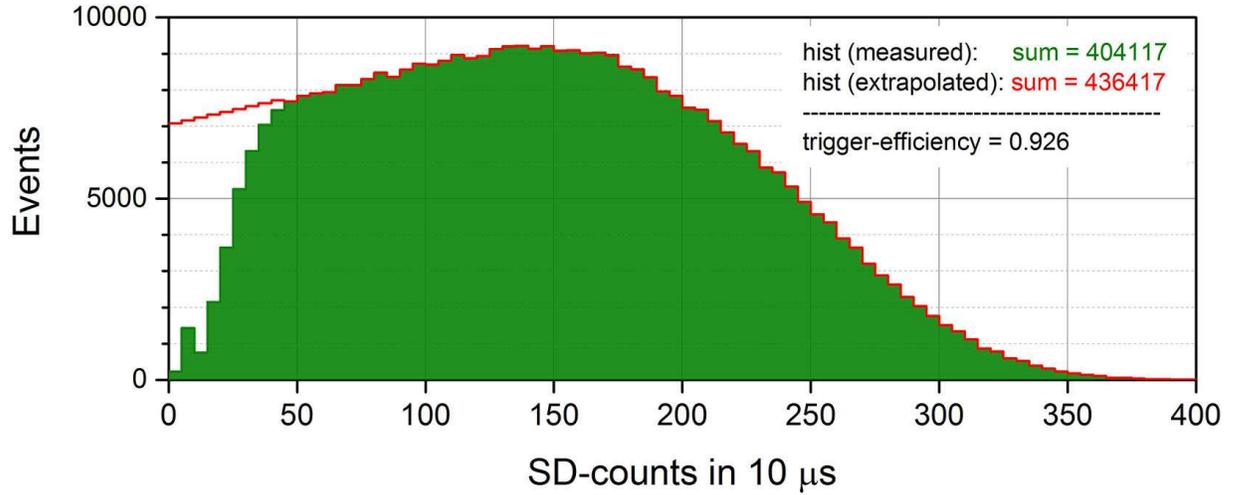}
\caption{
Calibration measurement to determine the neutron absorption rate in the detection unit.
The detection threshold is set as low as possible: gate(1)\,=\,80\,ns, N\,=\,4.
Other parameters for the signal processing are: gate(2$\ldots$N)\,=\,500\,ns,
b-time\,=\,100\,$\mu$s. The SiPM dark count rate is 100~kHz. Shown is the measured
histogram of the number of SD-counts within the first 10\,$\mu$s of the detected events
and extrapolation of this histogram to zero number of SD-counts.
The ratio of the number of events in the measured histogram to that in the extrapolated
one gives the trigger efficiency of 0.92. Taking into account the measured event rate
of 8.3\,Hz the neutron absorption rate is estimated to be 9.0\,Hz.
}
\end{figure}

\begin{figure}[b]
\centering
\includegraphics[width=1.0\columnwidth,clip]{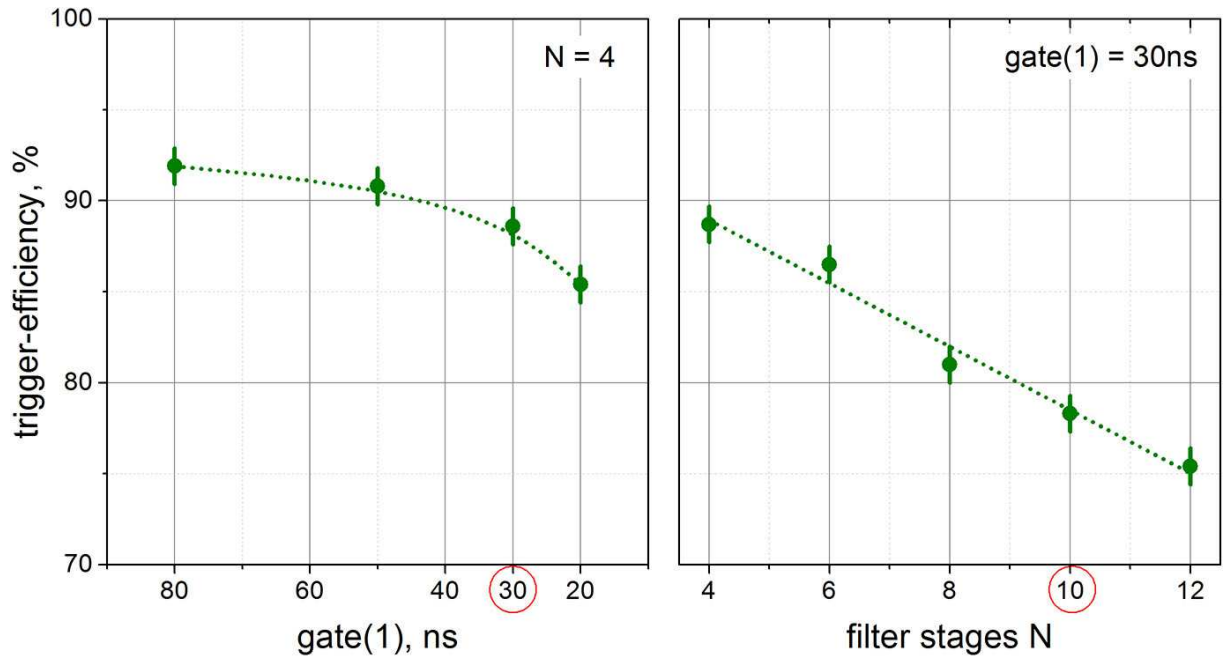}
\caption{
Trigger efficiency as a function of the gate width of the first filter stage gate(1)
measured at N\,=\,4 (left) and of the number of filter stages N measured
at gate(1)\,=\,30\,ns (right). Other parameters for the signal processing are fixed at:
gate(2$\ldots$N)\,=\,500\,ns, b-time\,=\,100\,$\mu$s. The dark count rate of the SiPM
is 100\,kHz. The chosen filter settings are indicated by red circles.
The dashed lines are shown to guide the eye.
}
\end{figure}

\clearpage
\newpage

\begin{figure}[t]
\centering
\includegraphics[width=0.75\columnwidth,clip]{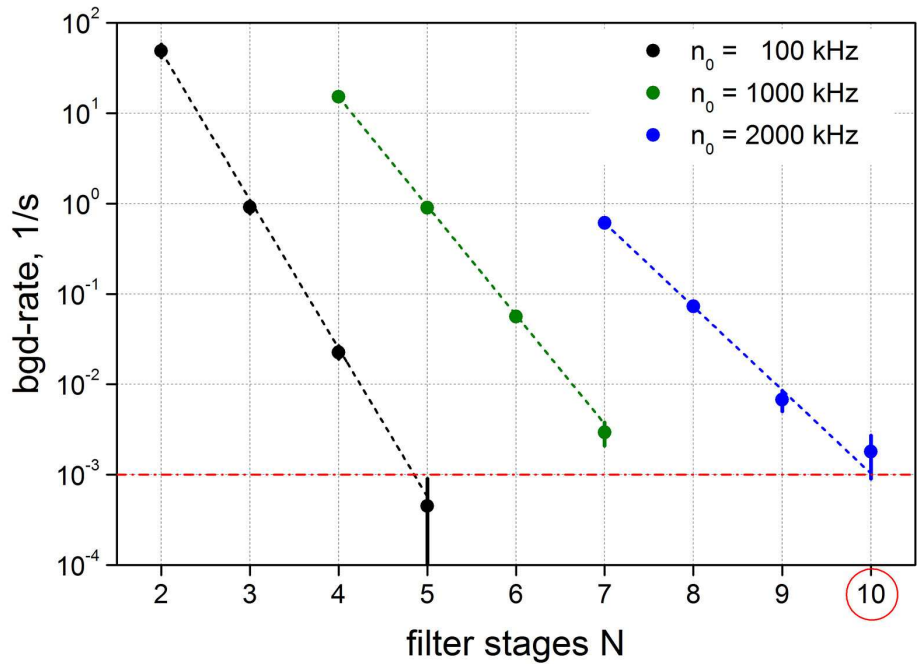}
\caption{
Background count rate of the detector as a function of the number of filter stages N
at different values of the SiPM dark count rate. Other parameters for
the signal processing are fixed at: gate(1)\,=\,30\,ns, gate(2$\ldots$N)\,=\,500\,ns,
b-time\,=\,100\,$\mu$s. The chosen filter setting is indicated by the red circle.
The dashed lines are shown to guide the eye.
}
\end{figure}

\begin{figure}[b]
\centering
\includegraphics[width=0.77\columnwidth,clip]{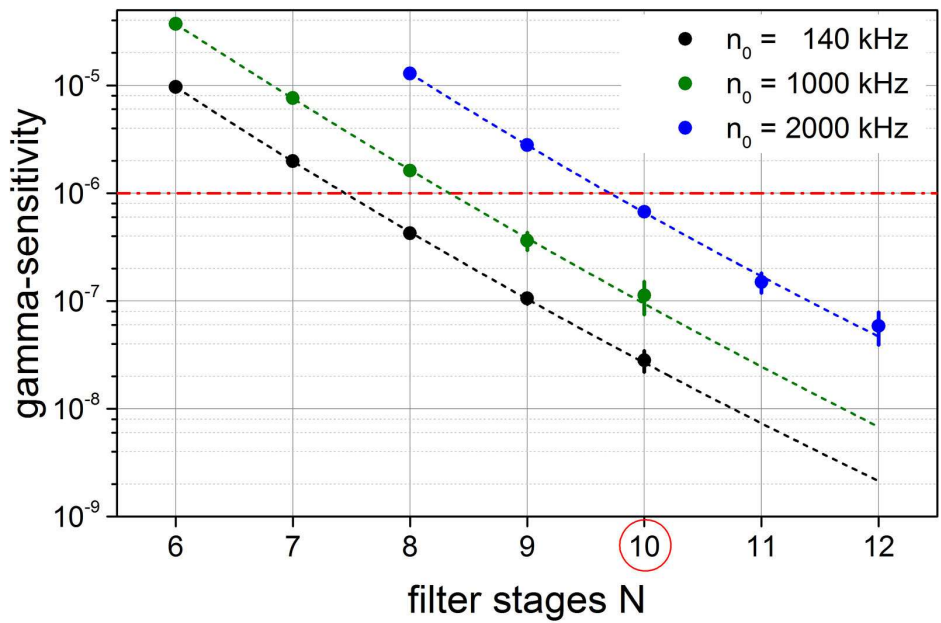}
\caption{
Gamma-sensitivity as a function of the number of filter stages N at different SiPM
dark count rates. Other parameters for the signal processing are fixed at:
gate(1)\,=\,30\,ns, gate(2$\ldots$N)\,=\,500\,ns, b-time\,=\,100\,$\mu$s.
The chosen filter setting is indicated by the red circle.
The dashed lines are shown to guide the eye.
}
\end{figure}

\clearpage
\newpage

\begin{figure}[t]
\centering
\includegraphics[width=0.8\columnwidth,clip]{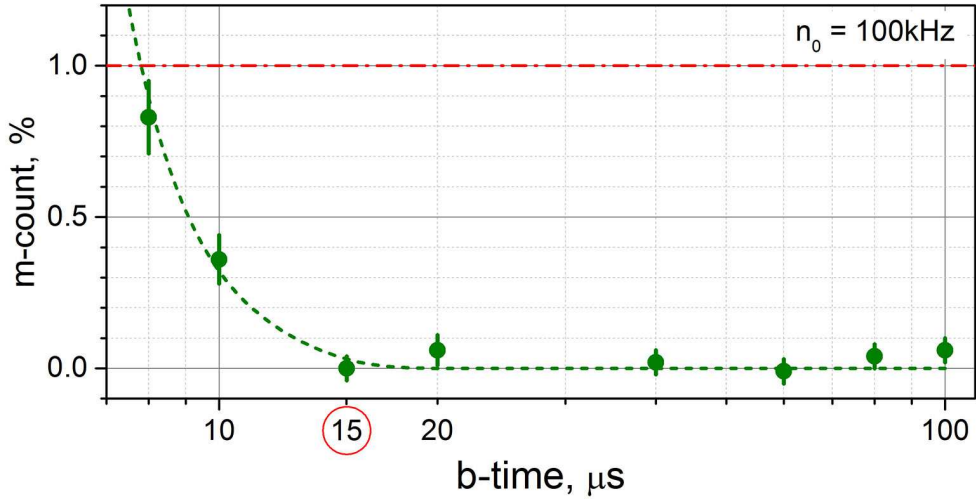}
\caption{
Multi-count ratio of the detector as a function of the set value of the blocking time
of the event generator. Other parameters for the signal processing are fixed at:
gate(1)\,=\,30\,ns, gate(2$\ldots$N)\,=\,500\,ns, N\,=\,10.
The dark count rate of the SiPM is 100\,kHz. The chosen b-time setting is indicated
by the red circle. The dashed line is shown to guide the eye.
}
\end{figure}

\begin{figure}[b]
\centering
\includegraphics[width=0.8\columnwidth,clip]{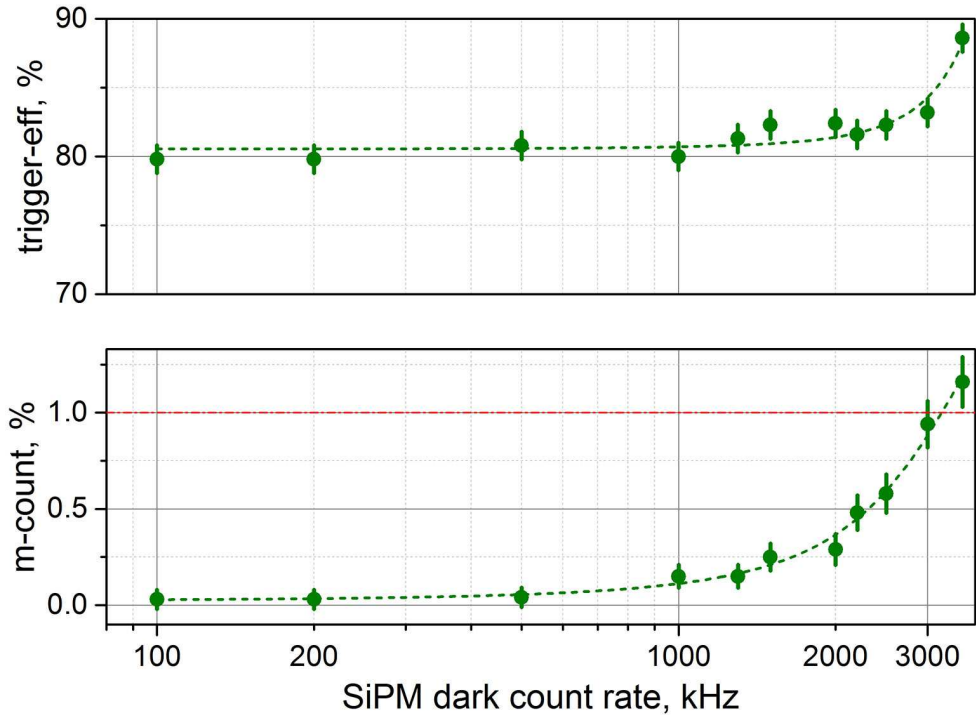}
\caption{
Trigger efficiency and multi-count ratio as a function of the dark count rate of the SiPM.
The parameters for the signal processing are fixed at: gate(1)\,=\,30\,ns,
gate(2$\ldots$N)\,=\,500\,ns, N\,=\,10, b-time\,=\,15\,$\mu$s.
The dashed lines are shown to guide the eye.
}
\end{figure}

\end{document}